\documentclass[prd,reprint,showpacs,nofootinbib]{revtex4-1}
\usepackage{array,graphicx,multirow,amsmath,amsfonts,mathrsfs,subfig}
\usepackage{nicefrac}
\usepackage{footmisc,xcolor}
\usepackage[T1]{fontenc}
\usepackage{tikz}
\usetikzlibrary{external}
\usetikzlibrary{tikzmark}

\newcommand{\re}[1]{(\ref{eq:#1})}

\def\phi{\varphi}

\def\rho{\varrho}
\def\d{\mathrm{d}}
\def\p{\partial}

\def\Im{\mathop{\rm Im}}

\renewcommand{\vec}[1]{\boldsymbol{#1}}

\newcommand{\rem}[1]{}

\def\={\discretionary{-}{-}{-}}

{\vskip 3pt plus 1pt minus 1pt\begin{normalsize}}%
{\par\end{normalsize}\vskip 3pt plus 1pt minus 1pt}

\newlength{\obrA} 
\newlength{\obrB} 

\newcommand{\thorn}{\text{\th}}
\newcommand{\eth}{\text{\dh}}

\newcommand{\kp}[1]{\psi_2^{\nicefrac{-#1}{3}}}

\newcommand{\REM}[1]{}
\newcommand{\Ka}{a}

\begin{document}

\title{Point particles and Appell's solutions on the axis of Kerr black hole for arbitrary spin in terms of the Debye potentials}

\date{\today}
\author{David Kofro\v{n}}
\email{d.kofron@gmail.com\\ In memoriam of Martin Scholtz.}
\affiliation{
Institute of Theoretical Physics, Faculty of Mathematics and Physics,\\
Charles University,\\
V Hole\v{s}ovi\v{c}k\'{a}ch 2, 180\,00 Prague 8, Czech Republic}


\begin{abstract}
The Teukolsky master equation --- a fundamental equation for test fields of any spin, or perturbations, in type D spacetimes --- is classically treated in its separated form. Then the solutions representing even the simplest sources -- point particles -- are expressed in terms of series. The only known exception is a static particle (charge or mass) in the vicinity of Schwarzschild black hole.

Here, we present a generalization of this result to a static point particle of arbitrary spin at the axis of Kerr black hole. A simple algebraic formula for the Debye potential from which all the NP components of the field under consideration can be generated is written down explicitly.

Later, we focus on the electromagnetic field and employ the classic Appell's trick (moving the source into a complex space) to get so called electromagnetic magic field on the Kerr background. Thus the field of nontrivial extended yet spatially bounded source is obtained.

We also show that a static electric point charge above the Kerr black hole induces, except an expected electric monopole, also a magnetic monopole charge on the black hole itself. This contribution has to be compensated. 

On a general level we discuss Teukolsky\,--\,Starobinsky identities in terms of the Debye potentials.

\end{abstract}
\maketitle

\section{Introduction}

Since the first exact solutions of the Einstein field equations were found, scientists became interested in behaviour of various test fields on these backgrounds. The very first successful attempt dating already back to the year 1928 was performed by Copson \cite{Copson1928} who discovered an electromagnetic potential of static electric charge in the vicinity of Schwarzschild black hole in a compact algebraic form considering Hadamard's theory of elementary solutions of partial differential equations.

The field of static electric charges later attracted a lot of attention in the seventies beginning with the work of Cohen and Wald in 1971 \cite{Cohen1971} (series expansion of electromagnetic field strength tensor). In 1972 Fackerell and Ipser \cite{Fackerell1972} made the first attempt to decouple the Maxwell equations in Newman\,--\,Penrose formalism, providing a separable equation for $\phi_{1}$ on the Schwarzschild background. A year later Teukolsky showed that on type~D background decoupled and separable equations for the outermost NP components of a test field of an arbitrary spin exist \cite{Teukolsky1973} providing an excellent framework for future progress. Solving the Maxwell equations in tensor components and series expansion Hanni and Ruffini \cite{Hanni1973} managed to draw lines of force of the field generated by a static point charge hovering above the Schwarzschild black hole.

Another breakthrough happened in 1974 when Cohen and Kegeles introduced the Debye potentials in the NP formalism \cite{Cohen1974b} --- from a single complex scalar field all the NP components of the field of arbitrary spin can be calculated.

Employing the Teukolsky equation the series expansion of the field of static charged point particle on the symmetry axis on the Kerr background was provided by Cohen, Kegeles and Wald \cite{Cohen1974}. Later a series expansion of fields of various electromagnetic sources was given by Bi\v{c}\'{a}k and Dvo\v{r}\'{a}k \cite{Bicak1976,Bicak1977} and independently by King \cite{King1976} in 1976\,--\,1977.

Yet the only known closed formula by Copson was rediscovered by Linet in \cite{Linet1976} 1976. 

The gravitational perturbations were treated in the Debye potential formalism in the works of Chrzanowski \cite{Chrzanowski1975} and Cohen and Kegeles \cite{Cohen1975,Kegeles1979} in 1975. The Copson formula was again rediscovered by Keidl, Friedman and Wiseman \cite{Keidl2007} in 2007 where they brought these results in the Debye potential formalism and generalized to static mass particle and gravitational perturbations. Later, these result were used by Sano and Tagoshi \cite{Sano2014} in 2014 in order to find a field of a massive disk around the Schwarzschild black hole.

In the section \ref{sec:meq} we recapitulate the Teukolsky master equation and the main results of the Debye potential formalism. We aslo present new results. Here we employ the NP formalism \cite{Newman1962} and its reformulation in terms of complex line bundles of $[p,q]$ weighted scalars --- GHP formalism \cite{Geroch1973}. 

The section \ref{sec:exact} is devoted to the exact solutions of test fields in non-separated form. IN the following section \ref{sec:elmag} we focus on the electromagnetic field. Visualisations of some results are provided and the Meissner effect, which has recently been proven to hold not only at the level of test fields on type D backgrounds but also in the non-perturbative mode and in general axially symmetric black hole spacetimes \cite{Gurlebeck2017,Gurlebeck2018}, is naturally observed.

\section{Master equation, Debye potentials and Teukolsky\,--\,Starobinsky identities}\label{sec:meq}
Teukolsky \cite{Teukolsky1973} provided decoupled equations for the NP components of gravitational perturbation $\dot{\psi_0}$ and $\dot{\psi_4}$, of test electromagnetic field $\phi_0$ and $\phi_2$ and neutrino field $\chi_0$ and $\chi_1$ in a general vacuum type D spacetime and performed a detailed analysis of these equations on the Kerr background. It has been realized that Rarita\,--\,Schwinger fields fit into the same scheme \cite{Guven1980}. 

Buchdahl \cite{Buchdahl1958,Buchdahl1962,Fordy1978} derived an algebraic constraint for solutions of minimally coupled spin $s\geq \nicefrac{3}{2}$ equation which is quite restrictive on curved background. 

In the spinor formalism \cite{Stewart1991} the zero rest mass equation for a test field of any spin $s$ is given in terms of a totally symmetric spinor $\phi_{AB\ldots C}$ of valence $2s$ as
\begin{equation}
		\nabla^{CC'}\phi_{AB\ldots C} = 0 \,,
		\label{eq:ZRM}
\end{equation}
and its solutions are restricted by the condition
\begin{equation}
		\Psi^{ABC}_{\phantom{ABC}(D}\,\phi_{EF\ldots )ABC}=0\,,
\end{equation}
where $\Psi_{ABCD}$ is the Weyl spinor. By projecting the equations \re{ZRM} onto the null tetrad and decoupling them one obtains equations similar to Teukolsky master equations \re{NPmin}\,--\,\re{NPmax} but \emph{without} the $(2s-1)(s-1)\psi_2$ term which represents coupling to the background curvature. This shows that the linearized gravity is not described by the spin-2 zero rest mass equation. Analogically the spin-$\nicefrac{3}{2}$ Rarita\,--\,Schwinger field is governed by equations obtained by linearization of supergravity field equations \cite{TorresdelCastillo1989}, not by the zero rest mass equation.

In the NP formalism \cite{Newman1962} quantities describing spacetime geometry and field equations are expressed in terms of scalars obtained as their projections onto the null tetrad $(\vec{l},\,\vec{m},\,\vec{\bar{m}}\,,\vec{n})$. The tetrad is determined by demanding that the only nonvanishing scalar products are $\vec{l}^a\vec{n}_a=-\vec{m}^a\vec{\bar{m}}_a=-1$.
The freedom in its choice is given by the Lorentz group which naturally splits into 4 groups: null rotations around fixed $\vec{l}$, null rotations around fixed $\vec{n}$, boosts in $(\vec{l},\,\vec{n})$ plane, and rotations in $(\vec{m},\,\vec{\bar{m}})$ plane. The metric is reconstructed by $\vec{g}_{ab}=-2\vec{l}_{(a}\vec{n}_{b)}+2\vec{m}_{(a}\vec{\bar{m}}_{b)}$.
And the connection is encoded in 12 complex spin coefficients. 

The discrete `prime' transformation 
\begin{align}
		\vec{l}&\stackrel{'}{\longleftrightarrow} \vec{n}\,, &
		\vec{m}&\stackrel{'}{\longleftrightarrow} \vec{\bar{m}}\,,
\end{align}
allows us to reduce the number of Greek letters needed for spin coefficients, and it is common to use $(\kappa,\,\sigma,\,\rho,\,\tau,\,\beta,\,\epsilon)$ and their primed counterparts.

In GHP formalism \cite{Geroch1973} the real null directions $\vec{l},\,\vec{n}$ are fixed and the freedom of the tetrad is restricted to boosts in $(\vec{l},\,\vec{n})$ plane and rotations in $(\vec{m},\,\vec{\bar{m}})$ plane which can be written explicitly as\footnote{This transformation naturally follows from the transformation of spin dyad $o^A\rightarrow \lambda o^A,\,\iota^A\rightarrow \lambda^{-1}\iota^A$.}
\begin{align}
		\vec{l}^a &\ \rightarrow\ \lambda\bar{\lambda}\, \vec{l}^a,\, &
		\vec{n}^a &\ \rightarrow\ \lambda^{-1}\bar{\lambda}^{-1}\,\vec{n}^a\,, \label{eq:boost}\\
		\vec{m}^a &\ \rightarrow\ \lambda\bar{\lambda}^{-1}\,\vec{m}^a,\, &
		\vec{\bar{m}}^a &\ \rightarrow\ \lambda^{-1}\bar{\lambda}\,\vec{\bar{m}}^a\,.\label{eq:rotation}
\end{align}
This allows us to define GHP scalar of a specific weight $[p,q]$ (corresponding to a spin- and boost-weight $(\frac{1}{2}(p-q),\,\frac{1}{2}(p+q))$) which transforms as 
\begin{equation}
		\phi \longrightarrow \lambda^p\bar{\lambda}^q\, \phi\,,
\end{equation}
under the transformations \re{boost}\,--\,\re{rotation}. The $(\kappa,\,\sigma,\,\rho,\,\tau)$\footnote{Together with their primed counterparts.\label{fn:prime}} are proper GHP scalars meanwhile neither $(\beta,\,\epsilon)$\footref{fn:prime} nor NP directional derivatives $(D,\,\Delta,\,\delta,\,\bar{\delta})$ transform properly. Incorporating $\beta$ and $\epsilon$ in differential operators leads to GHP derivatives
\begin{align}
		\thorn \eta &= \left(D-p\epsilon-q\bar{\epsilon}\right)\eta\,, &
		\thorn' \eta &= \left(\Delta+p\epsilon'+q\bar{\epsilon}'\right)\eta\,, \\
		\eth \eta &=\left(\delta-p\beta+q\bar{\beta}'\right)\eta \,, &
		\eth' \eta &=\left(\bar{\delta}+p\beta'-q\bar{\beta}\right)\eta \,,
\end{align}
which acting on scalar of weight $[p,q]$ create a scalar of weight $[p+r,q+s]$ where the appropriate raising/lowering weights $[r,s]$ of the particular derivative are as follows 
\begin{align}
		\thorn &\rightarrow [+1,+1]\,, &\thorn' &\rightarrow [-1,-1]\,, \\
		\eth &\rightarrow [+1,-1]\,,  &\eth' &\rightarrow [-1,+1]\,.
\end{align}
Therefore $\eth$ and $\eth'$ are spin raising and lowering operators, meanwhile $\thorn$ and $\thorn'$ are boost raisining and lowering operators.

The prime operation takes a scalar of weight $[p,q]$ into a scalar of weight $[-p,-q]$, and complex conjugation into a scalar of weight $[q,p]$.

The GHP formalism allows for a simple consistency test of equations: only a scalars of the same GHP weights can be compared. 

\begin{table}
\centering
\renewcommand{\arraystretch}{1.4}
\begin{tabular}{c@{\quad}|@{\quad}*{9}{c}}
$S$ &-2 & -$\nicefrac{3}{2}$& -1 & -$\nicefrac{1}{2}$& 0 & $\nicefrac{1}{2}$& 1 & $\nicefrac{3}{2}$& 2\\
$[p,q]$ &[-4,0] & [-3,0]& [-2,0]  & [-1,0] & [0,0] & [1,0] & [2,0] & [3,0]& [4,0] \\
\hline  
0 & & &  & & $\phi$ & & & & \\
$\nicefrac{1}{2}$ &&  & & $\chi_1$ & & $\chi_0$ & & & \\
1& & & $\phi_2$ & & \textcolor{gray}{$\phi_1$} & & $\phi_0$ & & \\
\nicefrac{3}{2}& & $\Sigma_3$ & & \textcolor{gray}{$\Sigma_2$} & & \textcolor{gray}{$\Sigma_1$} & & $\Sigma_0$ & \\
2& $\dot{\psi}_4$ & & \textcolor{gray}{$\dot{\psi}_3$} & &\textcolor{gray}{$\dot{\psi}_2$} & &\textcolor{gray}{$\dot{\psi}_1$} & &$\dot{\psi}_0$ 
\end{tabular}
\caption{The NP field components for field of a given spin $s$, their spin-weight $S$ in the first row, GHP weights $[p,q]$ in the second row. The outermost projections (with maximal/minimal spin-weight), for which the Teukolsky Master equation holds, are in black, the others in gray.}
\label{tab:fields}
\end{table}

Since we are interested in the NP components of test fields of an arbitrary spin $s$, and, moreover, we utilize the GHP formalism we summarize the notation in the Table~\ref{tab:fields}. 

The gravitational perturbations are described by 5 scalars $\dot{\psi}_j$ (meanwhile $\psi_2$ is the only non-zero component of the Weyl tensor of the background metric of type D), the Rarita\,--\,Schinger field by 4 scalars $\Sigma_j$, the electromagnetic field by 3 scalars $\phi_j$, the Weyl neutrino field by 2 scalars $\chi_j$ and the Klein\,--\,Gordon field by single scalar $\phi$. We denote the field components $\dot{\psi_0},\,\Sigma_0,\,\phi_0,\,\chi_0,\,\phi$ as $\Phi_{[2s,0]}$ and $\dot{\psi_4},\,\Sigma_3,\,\phi_2,\,\chi_1,\,\phi$ as $\Phi_{[-2s,0]}$ where in the square brackets are GHP weights $[p,q]$. In our notation the $s$ is strictly non-negative and just denotes the absolute value of the spin of the particular (outermost) component. In this way the equations for $\Phi_{[2s,0]}$ and $\Phi_{[-2s,0]}$ look similarly (which is the direct consequence of the prime ($'$) operation of GHP formalism) and we do not have to distinguish equations of positive and negative spin-weight $S$. In this notation the action of prime on $s$ is identity $s'=s$ (otherwise the prime would change the sign of $S$). The equations which are primed versions of the others will be grouped and labeled as sub\,--\,equations, such that $(a)'=(b)$.
\REM{
\begin{table*}
\centering
\caption{Notation, spin and GHP weight of field components.}
\renewcommand{\arraystretch}{1.5}
\begin{ruledtabular}
\begin{tabular}{l|ccccccccc} 
$\hat{\Phi}$ & 
$\Psi_4$ & $\Sigma^{RS}_3$ &
$\phi^{EM}_2$ & $\chi_1$ &
$\Phi^{KG}$ &
$\chi_0$ & $\phi^{EM}_0$ &
$\Sigma^{RS}_0$ & $\Psi_0$ \\
spin weight   &
-2 & -\nicefrac{3}{2} &
-1 & -\nicefrac{1}{2} &
0 &
\nicefrac{1}{2} & 1 &
\nicefrac{3}{2} & 2 \\
$s$    &
2 & \nicefrac{3}{2} &
1 & \nicefrac{1}{2} &
0 &
\nicefrac{1}{2} & 1 &
\nicefrac{3}{2} & 2 \\
GHP weight &
[-4,0] & [-3,0] & [-2,0] & [-1,0] & [0,0]& [1,0]& [2,0] & [3,0] & [4,0] \\
\end{tabular}
\end{ruledtabular}
\label{tab:F}
\end{table*}
}

The Teukolsky master equations in the NP formalism, as adapted from \cite{Teukolsky1973}, read
{\footnotesize
\begin{widetext}
\begin{eqnarray}
\bigl[ \left( D - (2s-1)\epsilon+\bar{\epsilon} -2s\varrho -\bar{\varrho} \right)\left( \Delta-2s\gamma+\mu \right) 
- \left( \delta +\bar{\pi}-\bar{\alpha}-(2s-1)\beta-2s\tau \right)\left( \bar{\delta}+\pi-2s\alpha \right)
-(2s-1)(s-1)\psi_2 \bigr] \Phi_{[2s,0]} &=& 0\label{eq:NPmin}\,, \\
\bigl[ \left( \Delta +(2s-1)\gamma-\bar{\gamma}+2s\mu+\bar{\mu} \right)\left( D+2s\epsilon-\varrho \right) 
- \left( \bar{\delta} -\bar{\tau}+\bar{\beta}+(2s-1)\alpha+2s\pi \right)\left( \delta-\tau+2s\beta \right)
 -(2s-1)(s-1)\psi_2 \bigr] \Phi_{[-2s,0]} &=& 0\,. \label{eq:NPmax} 
\end{eqnarray}
\end{widetext}}
Straightforward translation of these equations into the GHP formalism leads to
\begin{subequations}
\begin{eqnarray}
\bigl[ \left( \thorn -\bar{\rho}-2s\rho \right)\left( \thorn' -\rho' \right)  
- \left( \eth-\bar{\tau}'-2s\tau \right)\left( \eth'-\tau' \right)&&\label{eq:GHPmin}\\
 -(2s-1)(s-1)\psi_2 \bigr] \Phi_{[2s,0]} & =& 0   \,,\nonumber \\
\bigl[ \left( \thorn'-\bar{\rho}' -2s\rho' \right)\left( \thorn -\rho \right)  
- \left( \eth'-\bar{\tau}-2s\tau' \right)\left( \eth-\tau \right) &&\label{eq:GHPmax}\\ 
-(2s-1)(s-1)\psi_2 \bigr] \Phi_{[-2s,0]} & =& 0  \,.\nonumber
\end{eqnarray}
\end{subequations}
These equations are a cornerstone of the perturbation theory in the NP and GHP formalism.

\subsection{Teukolsky\,--\,Starobinsky identities}

It has been shown that there are relations in between $\dot{\psi}_0$ and $\dot{\psi}_4$ as well as in between $\phi_0$ and $\phi_2$, \cite{Starobinski1973,Teukolsky1974}. These relations, known as Teukolsky\,--\,Starobinsky identities, were intensively studied in their separated form \cite{Kalnins1989} in the Boyer\,--\,Lindquist coordinate chart, but they are truly geometrical identities which depend only on the special NP tetrad. Their covariant form for electromagnetic field read \cite{Aksteiner2019,Price} 
\begin{subequations}
\begin{eqnarray}
\thorn'\thorn'\left( \kp{2} \phi_0 \right)&=&\eth\eth\left( \kp{2} \phi_{2} \right) \,,\\
\thorn\thorn\left( \kp{2} \phi_{2} \right)&=&\eth'\eth'\left( \kp{2} \phi_0 \right) \,.
\end{eqnarray}
\end{subequations}

Meanwhile the Buchdahl constraint enforces the modification of the test field equations by introducing the coupling to the background curvature, for the gravitational field a new feature appears -- the backreaction of perturbations on the geometry, as can be seen from the additional term in the Teukolsky\,--\,Starobinsky identities for gravitational field:
\begin{subequations}
\begin{eqnarray}
\hspace{-2.5em}\thorn'\thorn'\thorn'\thorn' \left( \kp{4} \dot{\psi}_{0} \right) 
& = &
\eth\eth\eth\eth \left( \kp{4} \dot{\psi}_{4} \right) +3\mathcal{V}\,\bar{\dot{\psi}}_4 \,,\label{eq:TSIg0}\\
\thorn\thorn\thorn\thorn \left( \kp{4} \dot{\psi}_{4} \right)
& = &
\eth'\eth'\eth'\eth' \left( \kp{4} \dot{\psi}_{0} \right)  -3\mathcal{V}\,\bar{\dot{\psi}}_0 \label{eq:TSIg4}
\end{eqnarray}
\end{subequations}
where $\mathcal{V}$ is an operator defined by its action on scalar of GHP weight $[p,q]$
\begin{equation}
\mathcal{V} =\psi_2^{-\nicefrac{1}{3}}\left[ \tau'\eth-\tau\eth'+\rho\thorn'-\rho'\thorn+\frac{p}{2}\,\psi_2 +\frac{q}{2}\frac{\rho}{\bar{\rho}}\,\bar{\psi}_2\right]. 
\label{eq:}\end{equation} 
This is valid for non-accelerating vacuum type D spacetimes \cite{Price}. Clearly, $\mathcal{V}'=-\mathcal{V}$.

Summing up these results, we can write the Teukolsky\,--\,Starobinsky identities as ($g_2$ denoting the backreaction term for $s=2$ case)
\begin{subequations}
\begin{eqnarray}
\hspace{-2em}\thorn'{}^{\,2s}\left( \kp{2s} \Phi_{[2s,0]} \right)&=&
  \eth^{2s}\left( \kp{2s} \Phi_{[-2s,0]} \right) +g_2\,, \label{eq:TSIa}\\
\hspace{-2em}\thorn^{2s}\left( \kp{2s} \Phi_{[-2s,0]} \right)&=& 
  \eth'{}^{\,2s}\left( \kp{2s} \Phi_{[2s,0]} \right) +g_2'\,. \label{eq:TSIb}
\end{eqnarray}
\end{subequations}
For the sake of completeness let us write down also the third, independent, Teukolsky\,--\,Starobinsky identity for electromagnetic field, discovered in \cite{Coll1987,Aksteiner2019}
\begin{equation}
\left( \thorn\eth+\bar{\tau}'\thorn \right)\left( \kp{2}\phi_2 \right) = \left( \thorn'\eth'+\bar{\tau}\thorn' \right)\left( \kp{2}\phi_0 \right)\,, 
\label{eq:TSI3}
\end{equation}
which is identical with its primed version and also with its starred version (star operation being the other standard GHP permutation of the null tetrad, cf. \cite{Geroch1973}).

\subsection{The Debye potentials}
\begin{figure}[h!]
\centering
\newcommand{\DScale}{.38}
\newcommand{\DRadius}{.2}
\subfloat[$\phi_0,\,\phi_2$]{\ \includegraphics{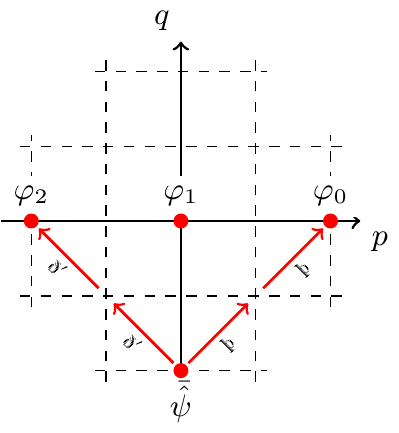}}\quad
\subfloat[$\phi_1$]{\ \includegraphics{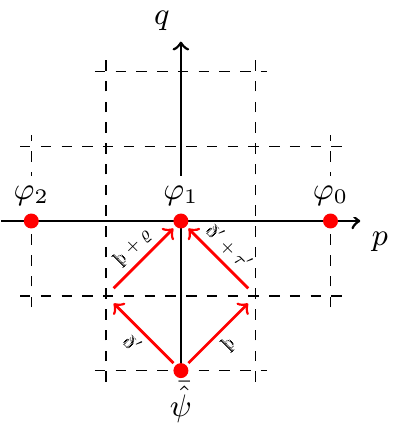}}
\caption{Electromagnetic NP scalars $\phi_0,\,\phi_1,\,\phi_2$ being obtained from the Debye potential $\bar{\hat{\psi}}_{[0,2]}$; $[p,q]$ diagram.}
\label{fig:DScheme}
\end{figure}

The Debye potential --- a single complex scalar field --- from which not only the spin-weight $+s$ and $-s$ NP components, but \emph{all} the field components, cf. $\dot{\psi}_0,\,\dot{\psi}_1,\,\dot{\psi}_2,\,\dot{\psi}_3,\,\dot{\psi}_4$, can be generated was introduced in \cite{Cohen1974a,Cohen1975}. There are more ways how to introduce this scalar potential \cite{Cohen1974a,Stewart1979} which is easily seen from the fact that in spinor formalism the Hertz potential is given by $\chi_{AB\dots C}=\hat{\psi}\, o_A o_B\dots o_C$ or its primed version $\chi_{AB\dots C}=\tilde{\psi}\, \iota_A \iota_B\dots \iota_C$ (in \cite{Cohen1974a} even the third possibility for the electromagnetic field, i.e. $\chi_{AB}=\check{\psi}\, o_{(A} \iota_{B)}$, is explicitly worked out). The NP projection of the equation for Hertz potential (see Eqs. (2.9), or (4.13) in \cite{Stewart1979}) are the equations for the Debye potential, given also sooner in \cite{Cohen1974a}, which we write here in GHP notation as follows
\begin{subequations}
\begin{eqnarray}
\bigl[ \left( \thorn -\bar{\rho} \right)\left( \thorn' +(2s-1)\rho' \right)  
- \left( \eth-\bar{\tau}'\right)\left( \eth'+(2s-1)\tau' \right) &&\label{eq:GHPDebT}\\
-(2s-1)(s-1)\psi_2 \bigr] \tilde{\psi}_{[2s,0]} & =& 0 \nonumber  \,,\\
\bigl[ \left( \thorn' -\bar{\rho}' \right)\left( \thorn +(2s-1)\rho \right)  
- \left( \eth'-\bar{\tau}\right)\left( \eth+(2s-1)\tau \right) &&\label{eq:GHPDeb}\\
-(2s-1)(s-1)\psi_2 \bigr] \hat{\psi}_{[-2s,0]} & =& 0  \,, \nonumber
\end{eqnarray}
\end{subequations}
while the test fields itself are generated by differentiation of the Debye potentials $\bar{\hat{\psi}}_{[0,-2s]}$ (or $\bar{\tilde{\psi}}_{[0,2s]}$, to be discussed later) as follows:
Klein\,--\,Gordon field
\begin{equation}
\phi=\bar{\hat{\psi}}_{[0,0]}\,,
\label{eq:}\end{equation}
neutrino field
\begin{eqnarray}
\chi_0 &=& \thorn\,\bar{\hat{\psi}}_{[0,-1]}\,, \\
\chi_1 &=& \eth'\,\bar{\hat{\psi}}_{[0,-1]}\,,
\end{eqnarray}
electromagnetic field
\begin{eqnarray}
\phi_{0} &=& \thorn\thorn\,\bar{\hat{\psi}}_{[0,-2]}\,, \label{eq:phi0}\\
2\phi_{1} &=& \left[ \left( \thorn+\rho \right)\eth'+\left( \eth'+\tau' \right)\thorn \right] \bar{\hat{\psi}}_{[0,-2]}\,, \label{eq:phi1}\\
 &=&2\left( \thorn\eth'+\tau'\thorn \right)\bar{\hat{\psi}}_{[0,-2]}\,, \nonumber \\
 &=&2\left( \eth'\thorn+\rho\eth' \right)\bar{\hat{\psi}}_{[0,-2]}\,, \nonumber \\
\phi_{2} &=& \eth'\eth'\, \bar{\hat{\psi}}_{[0,-2]}\,, \label{eq:phi2}
\end{eqnarray}
where the scheme of generating of the field components from the potential can be clearly seen and is visualised in Fig.~\ref{fig:DScheme}. There exists only one way how to get in two steps from $\bar{\hat{\psi}}$ to $\phi_0$ or $\phi_2$. On the contrary, there are two possibilities how to get to $\phi_1$ and both of them have to be taken into account. Later, of course, the expressions can be simplified using the commutators as in Eq. \re{phi1}.

We will not study the Rarita\,--\,Schwinger field in details, the reader can find the discussion in\footnote{This result is a consequence of their calculations, but it is not explicitly stated in \cite{TorresdelCastillo1989}.} \cite{TorresdelCastillo1989}, but let us note that
\begin{eqnarray}
\Sigma_0 &=& \thorn\thorn\thorn\, \bar{\hat{\psi}}_{[0,-3]}\,,\\
\Sigma_3 &=& \eth'\eth'\eth'\, \bar{\hat{\psi}}_{[0,-3]}\,.
\end{eqnarray}

The gravitational perturbations, which are becoming more complicated due the gauge dependence and the coupling to the background geometry, are not discussed here in their full complexity, the reader can find details in \cite{Kegeles1979,Stewart1979,Deadman2011}. The NP scalars perturbations\footnote{As in \cite{Kegeles1979} the NP scalars $\dot{\psi}_1,\,\dot{\psi}_2,\,\dot{\psi}_2$ are sums over all possible paths of length 4 connecting $\bar{\hat{\psi}}$ and $\dot{\psi}_j$ in $[p,q]$ diagram (operator ordering) and thus have $4,\,6,\,4$ summands, see \cite{Kegeles1979}. Simplifications by means of commutators are provided in \cite{Deadman2011,Aksteiner2019}.}  $\dot{\psi_0}$ and $\dot{\psi_4}$ in incoming radiation gauge read 
\begin{eqnarray}
2\dot{\psi}_0&=& \thorn\thorn\thorn\thorn\, \bar{\hat{\psi}}_{[0,-4]}\,,\label{eq:psi0}\\
2\dot{\psi}_4&=& \eth'\eth'\eth'\eth'\, \bar{\hat{\psi}}_{[0,-4]} \label{eq:psi4}\\
&& -3\psi_2\left[ \tau'\eth-\tau\eth'+\rho\thorn'-\rho'\thorn-2\psi_2 \right]\hat{\psi}_{[-4,0]}\nonumber\,,
\end{eqnarray}
where the backreaction exhibits itself in the second term in \re{psi4} for the gravitational perturbations, as discussed above the Eqs. \re{TSIg0}\,--\,\re{TSIg4}. 

It should be noted that in general (for arbitrary $s$)
\begin{equation}
		\eth'{}^{(2s)}\, \bar{\hat{\psi}}_{[0,-2s]}\,,
\end{equation}
as well as
\begin{align}
&\psi_2^{\nicefrac{2s}{3}} \psi_2^{-\nicefrac{1}{3}} 
	\left[ \tau'\eth-\tau\eth'+\rho\thorn'-\rho'\thorn- s\psi_2\right] \hat{\psi}_{[-2s,0]}\\
&\qquad\quad = \psi_2^{\nicefrac{2s}{3}} \mathcal{V}\, \hat{\psi}_{[-2s,0]}\,,
\end{align}
are linearly independent solutions of the Teukolsky master equation \re{GHPmax}.

From the Teukolsky\,--\,Starobinsky identities we observe, that the combination $\kp{2s}\Phi_{[\pm 2s,0]}$ seems to be somewhat fundamental. In \cite{Wald1978,Aksteiner2019} the Teukolsky master equation \re{GHPmin}, \re{GHPmax} was cast in the form where an operator acts on $\kp{2s}\Phi_{[\pm 2s,0]}$ instead of just plain $\Phi_{[\pm 2s,0]}$. The resulting equations, an equivalent form of \re{GHPmin} and \re{GHPmax}, read
\begin{subequations}
\begin{eqnarray}
\bigl[ \left( \thorn -\bar{\rho} \right)\left( \thorn' +(2s-1)\rho' \right)  
- \left( \eth-\bar{\tau}'\right)\left( \eth'+(2s-1)\tau' \right) &&\label{eq:GHPminR}\\ -(2s-1)(s-1)\psi_2 \bigr] \kp{2s} \Phi_{[2s,0]} & =& 0   \nonumber\,,\\
\bigl[ \left( \thorn' -\bar{\rho}' \right)\left( \thorn +(2s-1)\rho \right)  
- \left( \eth'-\bar{\tau}\right)\left( \eth+(2s-1)\tau \right)  &&\label{eq:GHPmaxR}\\-(2s-1)(s-1)\psi_2 \bigr] \kp{2s}\Phi_{[-2s,0]} & =& 0  \,. \nonumber
\end{eqnarray}
\end{subequations}

Clearly, the Eq. \re{GHPDeb} for the Debye potential $\hat{\psi}$ is identical with the Eq. \re{GHPmaxR} for the rescaled NP component of the field $\kp{2s}\Phi_{[-2s,0]}$ and the same equivalence holds for equations \re{GHPDebT} and \re{GHPminR}, see \cite{Wald1978}. 

In the next section we will show that similar relation holds in between the Fackerell\,--\,Ipser equation for $\phi_1$ and the third choice of the Debye potential for spin-1 field (i.e. $\check{\psi}o_{(A}\iota_{B)}$) .

From the equivalence of these equations, we immediately observe, that any solution of the Teukolsky master equation can serve as the Debye potential for the field of the same spin $s$
\begin{subequations}
\begin{align}
		\Phi_{[-2s,0]} \quad\longrightarrow\quad &&\hat{\psi}_{[-2s,0]}&=\psi_2^{-2s/3}\Phi_{[-2s,0]}\,,\\
		\Phi_{[2s,0]} \quad\longrightarrow\quad &&\tilde{\psi}_{[2s,0]}&=\psi_2^{-2s/3}\Phi_{[2s,0]}\,.
\end{align}
\end{subequations}
In this way  an infinite sequence of solutions can be generated from a single initial solution. 

\subsection{TSI in terms of the Debye potential}

In a compact notation we can write the field components in terms of the Debye potential for $s\leq \nicefrac{3}{2}$ as 
\begin{align}
\hat{\Phi}_{[2s,0]}  &= \thorn^{2s}\,\bar{\hat{\psi}}_{[0,-2s]}\,, &
\hat{\Phi}_{[-2s,0]} & = \eth'{}^{\,2s}\,\bar{\hat{\psi}}_{[0,-2s]}\,, \label{eq:TMEsolHat}
\end{align}
and $\Phi_{[4,0]}$ given by \re{psi0}, $\Phi_{[-4,0]}$ by \re{psi4}, which represents the solution of the Teukolsky master equation in term of the Debye potential $\hat{\psi}$. 

Substituting \re{TMEsolHat} into \re{TSIb} we can write this Teukolsky\,--\,Starobinsky identity as
\begin{equation}
\thorn{}^{\,2s}\left( \kp{2s}\eth'{}^{\,2s} \right)\bar{\hat{\psi}}_{[0,-2s]} =
\eth'{}^{\,2s}\left( \kp{2s}\thorn^{2s} \right)\bar{\hat{\psi}}_{[0,-2s]} \,,
\label{eq:TSIDebA}
\end{equation}
which holds even for gravitational perturbation since the backreaction contribution to $\dot{\psi}_4$ given by $-3\psi_2^{\nicefrac{4}{3}}\mathcal{V}\,\hat{\psi}_{[-4,0]}$ exactly cancels out  the term $-3\mathcal{V}\,\bar{\dot{\psi}}_0$ when expressed in term of the Debye potentials (utilizing the fact that the operator $\mathcal{V}$ commutes with all the GHP derivatives by construction). 

Explicitly performing the commutators in \re{TSIDebA}, for $s\in\left( \nicefrac{1}{2},\,1,\, \nicefrac{3}{2},\,2 \right)$ it is proven that this commutator is identically zero; $\zeta_{[0,-2s]}$ is now an arbitrary complex scalar function which does not have to be the solution of the Debye potential equation
\begin{equation}
\left[ \kp{2s} \thorn^{2s},\, \kp{2s} \eth'{}^{2s} \right]\zeta_{[0,-2s]} = 0\,,
\end{equation}
which is a manifestation of underlying symmetry of type D spacetimes.

On the other hand substituting \re{TMEsolHat} into the second Teukolsky\,--\,Starobinsky identity \re{TSIa} leads to 
\begin{equation}
\thorn'{}^{\,2s}\left( \kp{2s}\thorn^{2s} \right)\bar{\hat{\psi}}_{[0,-2s]} =
\eth^{2s}\left( \kp{2s}\eth'{}^{\,2s} \right)\bar{\hat{\psi}}_{[0,-2s]}\,,\label{eq:TSIDebb}
\end{equation}
only for $s\leq \nicefrac{3}{2}$ and for $s=2$ we get
\begin{multline}
\thorn'{}^{\,4}\left( \kp{4}\thorn^{4} \right)\bar{\hat{\psi}}_{[0,-4]} =\\
\eth^{4}\left( \kp{4}\eth'{}^{\,4} \right)\bar{\hat{\psi}}_{[0,-4]}-9\mathcal{V}\bar{\psi}_2^{\nicefrac{4}{3}}\mathcal{\bar{V}}\bar{\hat{\psi}}_{[0,-4]}\,.\label{eq:TSIDebB}
\end{multline}
These identities hold only if $\hat{\psi}$ is the solution of \re{GHPDeb}. The apparent asymmetry (or broken symmetry of prime operation in between \re{TSIg0} and \re{TSIg4}) of the Teukolsky\,--\,Starobinsky identities in terms of the Debye potentials follow from the fact that under the prime operation also the Hertz potential transforms as $\hat{\psi}o_A o_B o_C o_D \rightarrow \tilde{\psi} \iota_A \iota_B \iota_C \iota_D$ and the expression of $\dot{\psi}_0$ and $\dot{\psi}_4$ also breaks the symmetry.

Using the Debye potential $\tilde{\psi}$ we can draw the same conclusions for GHP scalars of weight $[0,2s]$, moreover their primed versions hold too.

\section{Point particles}\label{sec:exact}
\subsection{Preliminary --- Kerr black hole}
The rotating black hole --- the Kerr solution --- was discovered in 1963 by Roy Kerr \cite{Kerr1963}. Recent historical reviews on this fundamental solution can be found in \cite{KerrHistory,Teukolsky2015}.

Although the Kerr solution is well known we feel obliged to summarize the metric and the NP tetrad which we use, so that the further presented results are self-contained.

The metric itself in Boyer\,--\,Lindquist coordinates reads
\begin{multline}
\d s^2 = -\frac{\Delta}{\Sigma} \left( \d t -\Ka\sin^2\theta\,\d\phi \right)^2 +\frac{\Sigma}{\Delta}\,\d r^2 + \Sigma \, \d\theta^2  \\
+\frac{\sin^2\theta}{\Sigma}\left( \left( \Ka^2+r^2 \right)\d\phi - \Ka\,\d t \right)^2\,,
\label{eq:KerrMetric}
\end{multline}
with the standard definitions $\Delta=r^2-2Mr+\Ka^2$ and $\Sigma=r^2+\Ka^2\cos^2\theta$ where $M$ is the mass of a black hole and $Ma$ is its angular momentum.

The Kinnersley NP\footnote{Notice the boost given by $\sqrt{2}$ in contrast to standard textbook form. This makes the resulting expressions in term of the Debye potentials to appear `more symmetrical'.} tetrad $(\vec{l},\,\vec{m},\,\bar{\vec{m}},\,\vec{n})$ adapted to the principal null directions of the Weyl tensor read as follows
\begin{equation}
\begin{alignedat}{1}
\vec{l} &= \frac{1}{\sqrt{2}\,\Delta}\left[ \left( r^2+\Ka^2 \right)\vec{\p_t}+\Delta\,\vec{\p_r} + \Ka\,\vec{\p_\phi} \right]\,, \\
\vec{n} &= \frac{1}{\sqrt{2}\,\Sigma}\left[ \left( r^2+\Ka^2 \right)\vec{\p_t}-\Delta\,\vec{\p_r} + \Ka\,\vec{\p_\phi} \right]\,, \\
\vec{m} &= \frac{1}{\sqrt{2}\,\left( r+i\Ka\cos\theta \right)}\bigl( i\Ka\sin\theta\,\vec{\p_t}+\vec{\p_\theta} \\
&\qquad+i\csc\theta\,\vec{\p_\phi} \bigr)\,.
\end{alignedat}
\label{eq:NPtetrad}
\end{equation}
The nonzero NP spin coefficients are listed below
\newcommand{\Krho}{\left(r-i \Ka\cos\theta\right)}
\newcommand{\Krhocc}{\left(r+i \Ka\cos\theta\right)}
\begin{equation}
\begin{alignedat}{2}
\pi &= \frac{i}{\sqrt{2}}\,\frac{\Ka \sin\theta}{\Krho^2}\,,\quad &
\mu &= \frac{-1}{\sqrt{2}}\,\frac{\Delta}{\Sigma\Krho}\,, \\
\tau &= \frac{-i}{\sqrt{2}}\,\frac{\Ka \sin\theta}{\Sigma}\,, &
\rho &= \frac{-1}{\sqrt{2}}\,\frac{1}{\Krho}\,,\\
\gamma &=\mu +\frac{1}{\sqrt{2}}\,\frac{r-M}{\Sigma}\,, &
\beta &= \frac{1}{\sqrt{2}}\,\frac{\cot \theta}{\Krhocc}\,, \\
\alpha &= \pi-\bar{\beta}\,, & 
\end{alignedat}\label{eq:spinc}
\end{equation}
and the only nonzero Weyl scalar reads 
\begin{equation}
\psi_2 = -\frac{M}{\Krho^3} \,.
\label{eq:psi2}\end{equation}

\subsection{Exact Debye potentials}

\newcommand{\rmm}{\tilde{r}}
\newcommand{\am}{\tilde{r}_0}
\newcommand{\az}{r_0}

In this section we will solve the Debye potentials for point particles, generalizing known results (a) to an arbitrary position in the case of the Schwarzschild background or (b) to a particle on the rotation axis for the Kerr background.
 
\subsubsection{(a) Arbitrary position, non-rotating background}

Let us recall that Copson found and Linet later corrected solution for a static point charge on the Schwarzschild background in terms of static potential $V(r,\,\theta,\,\phi)$, i.e. the component of four potential $\vec{A}=V(r,\,\theta,\,\phi)\,\vec{\d} t$, as
\begin{equation}
V=\frac{e\left( \rmm\am-M^2\upsilon \right)}{\az r\sqrt{\rmm^2+\am^2-2\rmm\am\upsilon-M^2\left( 1-\upsilon^2 \right)}}+\frac{em}{\az r}\,, \label{eq:Copson}
\end{equation}
where location of the particle is encoded in its radial position $\az$ and angular coordinates $\theta_0,\,\phi_0$, which are hidden in the function 
\begin{equation}
\upsilon (\theta,\,\phi) = \cos\theta \cos\theta_0+\sin\theta \sin\theta_0 \cos\left( \phi-\phi_0 \right)\,.
\end{equation}
For the sake of clarity we also introduced shortcuts
\begin{align}
\rmm &= r-M \,, & \am &=\az-M\,, \label{eq:rmam}
\end{align}
which enter the equations frequently, and define
\begin{equation}
\xi = \rmm^2+\am^2-2\, \rmm\, \am\,\upsilon-M^2\left( 1-\upsilon^2 \right)\,,
\label{eq:dist}\end{equation}
which is a square of generalized distance from the point particle to the field point (cf. law of cosines).

In \cite{Keidl2007} the Debye potential for point charge ($s=1$) and point-like particle ($s=2$) on the (coordinate) axis of the Schwarzschild background was given.

Here, we provide the solution representing static point particle of any spin at arbitrary position in the Schwarzschild spacetime. The Teukolsky master equation \re{NPmin} is written for 
\begin{equation}
\Phi_{[2s,0]} = \frac{w(r,\,\theta,\,\phi)}{\sin^{s}\theta}\,,
\label{eq:TMEminSchwAn}
\end{equation}
and reads as follows
\begin{equation}
\begin{alignedat}{1}
&\frac{\left( \Delta^{1+s} w_{,r} \right)_{,r}}{\Delta^s} 
+\frac{\left( \sin^{1-2s}\theta\, w_{,\theta} \right)_{,\theta}}{\sin^{1-2s}\theta}\\
&\qquad\qquad+\frac{2is\cos\theta\, w_{,\phi}+w_{,\phi\phi}}{\sin^2\theta}+2sw = 0\,.
\end{alignedat}
\label{eq:TMEminSchw}
\end{equation}
The solutions are
\begin{equation}
\Phi_{[2s,0]} = \frac{ \left( \upsilon_{,\theta}\sin\theta + i\upsilon_{,\phi} \right)^s}{\xi^{s+\nicefrac{1}{2}}\,\sin^s\theta} \,.
\label{eq:TMEsol}\end{equation}

The equation for the Debye potential reads, for $\bar{\hat{\psi}}=W(r,\,\theta,\,\phi)/\sin^s\theta$, as follows
\begin{equation}
\begin{alignedat}{1}
&\Delta^s \left( \Delta^{1-s} W_{,r} \right)_{,r}
+\frac{\left( \sin^{1-2s}\theta\, W_{,\theta} \right)_{,\theta}}{\sin^{1-2s}\theta}\\
&\qquad\qquad\qquad\qquad +\frac{2is\cos\theta\, W_{,\phi}+W_{,\phi\phi}}{\sin^2\theta} = 0\,.
\end{alignedat}
\label{eq:DedminSchw}
\end{equation} 
The Debye potentials corresponding to the solutions \re{TMEsol} can be obtained by integration, since we have
\begin{equation}
\Phi_{[2s,0]}= \thorn^{2s}\,\bar{\hat{\psi}}_{[0,-2s]} = \left( \frac{1}{\sqrt{2}}\frac{\p}{\p r} \right)^{2s}\bar{\hat{\psi}}\,.
\label{eq:PiDSchw}
\end{equation}
It is a well known fact, that the lowest multipole part of the field of $l\leq s$ are not treated by the Teukolsky master equation, since the radiative NP components are completely insensitive to these modes. But these terms are included in the Debye formalism. In \cite{Keidl2007} they found that to obtain all the NP field components they have to include so called ``homogeneous'' solutions of the equations for the Debye potentials. These ``homogeneous'' solutions naturally arose as integration constants while solving the equation \re{PiDSchw}. We will discuss these terms later in the general Kerr background.

For now let us present the resulting Debye potential which is, in general, for arbitrary spin (except half-integers)
\begin{equation}
\bar{\hat{\psi}}_{[0,-2s]} = \frac{ \xi^{s-\nicefrac{1}{2}}\left( \upsilon_{,\theta}\sin\theta + i\upsilon_{,\phi} \right)^s}{\left( 1-\upsilon^2 \right)^s\,\sin^s\theta} \,.
\label{eq:DebPotSchw}
\end{equation}
For the sake of completeness let us mention that the integration of \re{PiDSchw} is straightforward and lead to 
\begin{multline}
\bar{\hat{\psi}}_{[0,-1]} = \frac{\sqrt{\upsilon_{,\theta}\sin\theta + i\upsilon_{,\phi}}}{\sqrt{\sin\theta}} \times \\
\frac{\arctan\left(\frac{\rmm-\am\upsilon}{\sqrt{(\am+M)(\am-M)(1-\upsilon^2)}}\right)}{\sqrt{(\am+M)(\am-M)(1-\upsilon^2)}}\,,
\label{eq:}\end{multline}
for $s=\nicefrac{1}{2}$ and to little a bit longer expressions for $s=\nicefrac{3}{2}$ which we do not present here.

These results can be used for calculating field of more complicated sources, since we know the Green function of the Debye potential equation.

\subsubsection{(b) On the axis of the Kerr black hole}
The next possible generalization is the point particle located on the symmetry axis of the Kerr black hole. The Teukolsky master equation \re{NPmin}, written for static axially symmetric field, for ansatz as in \re{TMEminSchwAn} reads as follows 
\begin{equation}
\frac{\left( \Delta^{1+s} w_{,r} \right)_{,r}}{\Delta^s} 
+\frac{\left( \sin^{1-2s}\theta\, w_{,\theta} \right)_{,\theta}}{\sin^{1-2s}\theta} +2sw = 0\,.
\label{eq:TMEminKerr}\end{equation}
The solutions representing point particles of arbitrary spin located on the axis at $r=\az$ are 
\begin{equation}
\Phi_{[2s,0]} = \frac{\sin^s\theta}{\xi_\Ka^{s+\nicefrac{1}{2}}} \,,
\label{eq:TMEsolK}\end{equation}
where now $\xi_\Ka$ is
\begin{equation}
\xi_\Ka = \rmm^2+\am^2-2\,\rmm\,\am\cos\theta-\left(M^2-\Ka^2\right)\sin^2\theta\,,
\label{eq:distK}\end{equation}
with $\rmm$ and $\am$ as in \re{rmam}. 
The equation for the Debye potential for $\bar{\hat{\psi}}=W/\sin^s\theta$ read 
\begin{equation}
\Delta^s \left( \Delta^{1-s} W_{,r} \right)_{,r}
+\frac{\left( \sin^{1-2s}\theta\, W_{,\theta} \right)_{,\theta}}{\sin^{1-2s}\theta}
 = 0\,,
\label{eq:DebminK}
\end{equation} 
with the solutions for non half-integer spin as follows
\begin{equation}
\bar{\hat{\psi}} = \frac{\xi_\Ka^{s-\nicefrac{1}{2}}}{\sin^s\theta}\,,
\label{eq:DebyeK}\end{equation}
and we do not present the results of integration for half-integer spins here.

At this point it is worth to discuss the $l<s$ contributions to the field which we have omitted so far because we dealt with solutions in compact, non-separated, form.

The equation for the Debye potential is separable and for 
\begin{equation}
\bar{\hat{\psi}} = e^{-i\omega t} R(r) S(\theta) e^{im\phi}\,,
\end{equation}
we get following set of ordinary differential equations
\begin{eqnarray}
\frac{\Delta^s \left( \Delta^{1-s} R_{,r} \right)_{,r}}{R} + \frac{K^2+2is\left( r-M \right)K}{\Delta}\quad && \\
-4is\omega r - \left(\Ka \omega-m\right)^2+m^2 -A  &=& 0\,,\nonumber\\
\frac{\left( \sin\theta\ S_{,\theta} \right)_{,\theta}}{\sin\theta\ S} + \left( \Ka\omega\cos\theta-s \right)^2 \qquad\qquad\qquad\qquad &&\\ 
-\frac{\left( m+s\cos\theta \right)^2}{\sin\theta} -s(s+1)+A &=&0\,,\nonumber
\end{eqnarray}
which are equivalent to the Teukolsky equations \cite{Teukolsky1973} up to sign change for spin and where $K=\left(\Ka^2+r^2\right)\omega-\Ka m$ as usually.

We will restrict ourselves to the static axially symmetric configurations and using these equations we can classify the $l<s$ contributions to the field using $l$.

For $s=1$ and $l=0$ the general solutions are
\begin{align}
R(r) &= A_1 r + A_0 \,,\\
S(\theta) &= B_1 \,\frac{\cos\theta}{\sin\theta} + B_{0}\, \frac{1}{\sin\theta} \,.
\label{eq:Spin1}
\end{align}
From the general solution $\bar{\hat{\psi}}=R(r)S(\theta)$ the monopole electromagnetic field is generated by equations \re{phi0}\,--\re{phi2}, namely
\begin{align}
\phi_0 &=0\,,\\
2\phi_1 &=\frac{A_0B_1-i\Ka A_1B_0}{\left( r-i\Ka\cos\theta \right)^2}\,, \label{eq:phi1mon}\\
\phi_2 &=0\,,
\label{}
\end{align}
which we will discuss in the next Section. Let us just note that the solution 
\begin{equation}
A_0 B_0 \,\frac{1}{\sin\theta} + A_1 B_1 \,r\,\frac{\cos\theta}{\sin\theta}
\label{eq:}\end{equation}
do not contribute to the physical field at all and thus are just calibration fields.

For $s=2$ and $l=0$ we get
\begin{align}
R(r) &= A_1\left( r-M \right)+A_2 \left( r^2-\Ka^2 \right)\,,\\
S(\theta) &= B_1 \frac{\cos\theta}{\sin^2\theta} + B_2\frac{\cos^2\theta+1}{\sin^2\theta}\,,
\label{}
\end{align}
and for $s=2$, $l=1$ the solution is as follows
\begin{align}
R(r) &= A_0 + A_3\left( \frac{1}{3}r^3-M r^2+\Ka^2 r \right)\,,\\
S(\theta) &= B_0\,\frac{1}{\sin^2\theta}+B_3\,\frac{\left( \cos^2\theta-3 \right)\cos\theta}{\sin^2\theta}\,,
\label{}
\end{align}
Let us just note that there is a systematics --- the multipole structure --- in these contribution (which has not been recognized in \cite{Keidl2007}).  

\section{Electromagnetic field, a deeper analysis}\label{sec:elmag}
Let us recall that for electromagnetic NP scalar $\phi_1$ a decoupled, albeit in general non-separable, equation was found by Fackerell and Ipser \cite{Fackerell1972}. In the GHP formalism (and for $\phi_1$ rescaled by $\kp{2}$ again) we can write it down as
\begin{equation}
\left[ \left( \thorn'+\rho'-\bar{\rho}' \right)\thorn - \left( \eth'-\bar{\tau}+\tau' \right)\eth \right]\kp{2}\phi_1 = 0\,.
\label{eq:FI}\end{equation}
Surprisingly, adapting the equation for the third possible choice of the Debye potential (Eq. (3.34) of \cite{Kegeles1979}) to the GHP formulation we get
\begin{equation}
		\left[ \left( \thorn'+\rho'-\bar{\rho}' \right)\thorn - \left( \eth'-\bar{\tau}+\tau' \right)\eth \right]\check{\psi}_{[0,0]} = 0\,,
\end{equation}
which is identical to Eq. \re{FI} and invariant under prime operation (can be checked using GHP commutators). Since \cite{Kegeles1979} contains typos in the expressions of the electromagnetic NP scalars, and we found some simplifications, let us present here the correct formulas
\begin{align}
\phi_0 &= 2\left( \thorn -\bar{\rho} \right)\eth\,\bar{\check{\psi}}\,,\\
\phi_1 &= \left[\left( \thorn+\rho-\bar{\rho} \right)\thorn' +\left( \eth'+\tau'-\bar{\tau} \right)\eth \right]\,\bar{\check{\psi}}\,,\label{eq:p11}\\
 &=2\left[ \eth'\eth+\left( \rho-\bar{\rho} \right)\thorn' \right]\bar{\check{\psi}}\,,\nonumber\\
\phi_2 &= 2\left( \thorn' -\bar{\rho}' \right)\eth'\,\bar{\check{\psi}}\,.
\label{<+label+>}
\end{align}
Similar scheme as Fig.~\ref{fig:DScheme} can be produced. Now there are two possible ways to get (in $[p,q]$ diagram) to the points  $\phi_0$ and $\phi_2$ and even four possible ways to $\phi_1$. The presented results are already simplified using commutators and also the equation for the Debye potential.

\vspace{1ex}
In the following calculations we will use the Debye potential $\hat{\psi}$, since it is defined for an arbitrary spin and the resulting expressions are more compact and useful (we have calculated $\check{\psi}$ for a point charge, but it is not worth publishing).

Using the equation for the Debye potential $\hat{\psi}$ for simplification of the results the static axially symmetric electromagnetic field reads as follows
\begin{align}
2\phi_0 &=\bar{\hat{\psi}}_{,rr}\,,\\
2\phi_1 &=\left( \frac{\bar{\hat{\psi}}_{,\theta}}{r-i\Ka\cos\theta} \right)_{,\theta}-\frac{i\left( ir\cos\theta+\Ka \right)\bar{\hat{\psi}}_{,r}+\cos\theta\, \bar{\hat{\psi}}}{\left( r-i\Ka\cos\theta \right)^2\sin\theta}\,,\\
2\phi_2 &=\frac{\Delta}{\left( r-i\Ka\cos\theta \right)^2}\,\bar{\hat{\psi}}_{,rr} \,.
\label{}
\end{align}

Total electric charge $Q_e$ and magnetic charge $Q_m$ can be calculated by integrating twoform $\vec{F}^*=\vec{F}-i\star\vec{F}$ over a closed 2-surface. This yields
\begin{equation}
iQ_e+Q_m = \frac{1}{4\pi} \oint \vec{F}^*\,.
\label{eq:}\end{equation}
After standard reconstruction of $\vec{F}^*$ from the NP components and the NP tetrad we get for surfaces of constant $t$ and $r$  
\begin{multline}
iQ_e+Q_m = 2\pi\int_0^\pi -\left( r-i\Ka\cos\theta \right)\Ka\sin^2\theta \,\phi_2 \\
-2i\sin\theta \left( r^2+\Ka^2 \right) \phi_1  
+\frac{\Ka\Delta\sin^2\theta}{r-i\Ka\cos\theta}\,\phi_0
\ \d\theta \,.
\label{eq:totQ}\end{multline}
\begin{figure*}[t!]
\centering
\subfloat[$\Ka=1$]{\includegraphics[width=0.32\obrA,keepaspectratio]{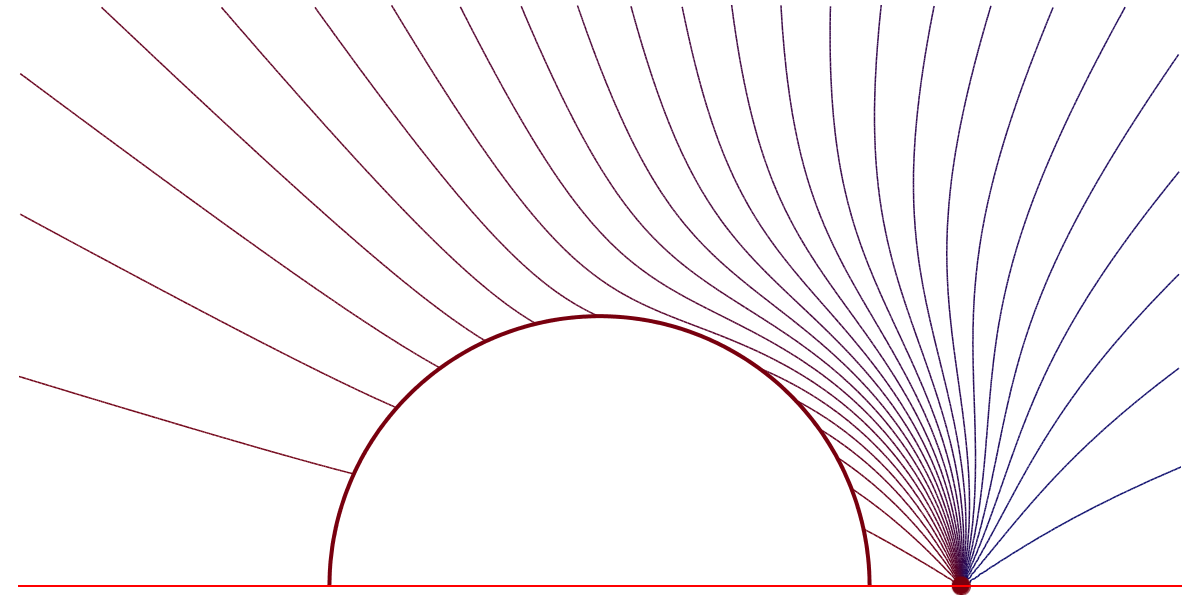}}\qquad
\subfloat[$\Ka=1.9$]{\includegraphics[width=0.32\obrA,keepaspectratio]{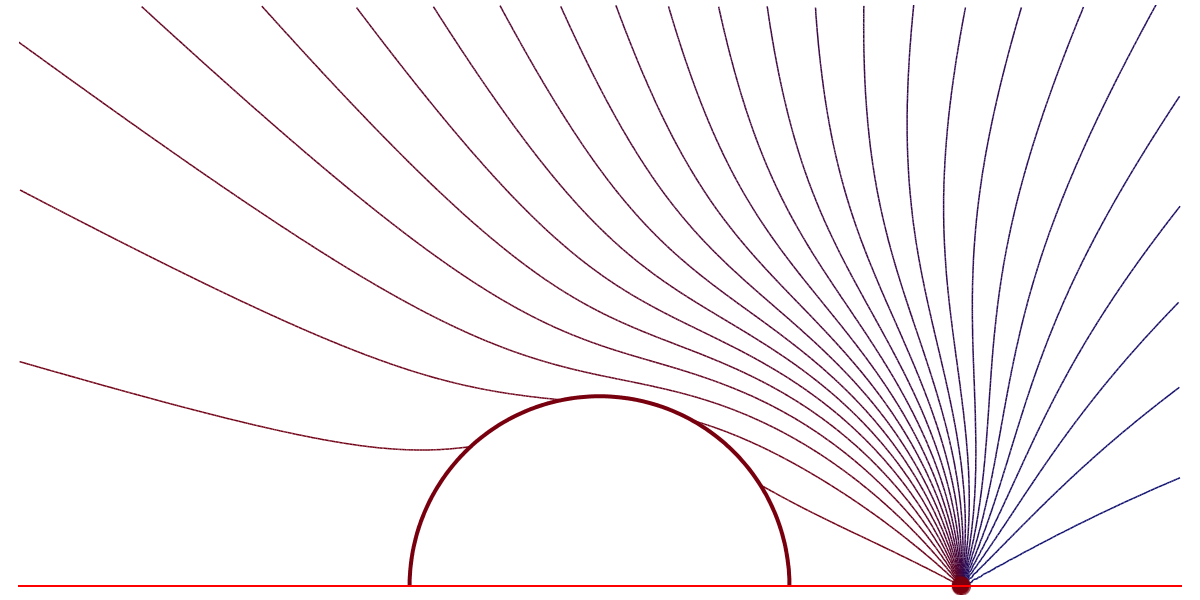}}\qquad
\subfloat[$\Ka=2$ (extremal Kerr)]{\includegraphics[width=0.32\obrA,keepaspectratio]{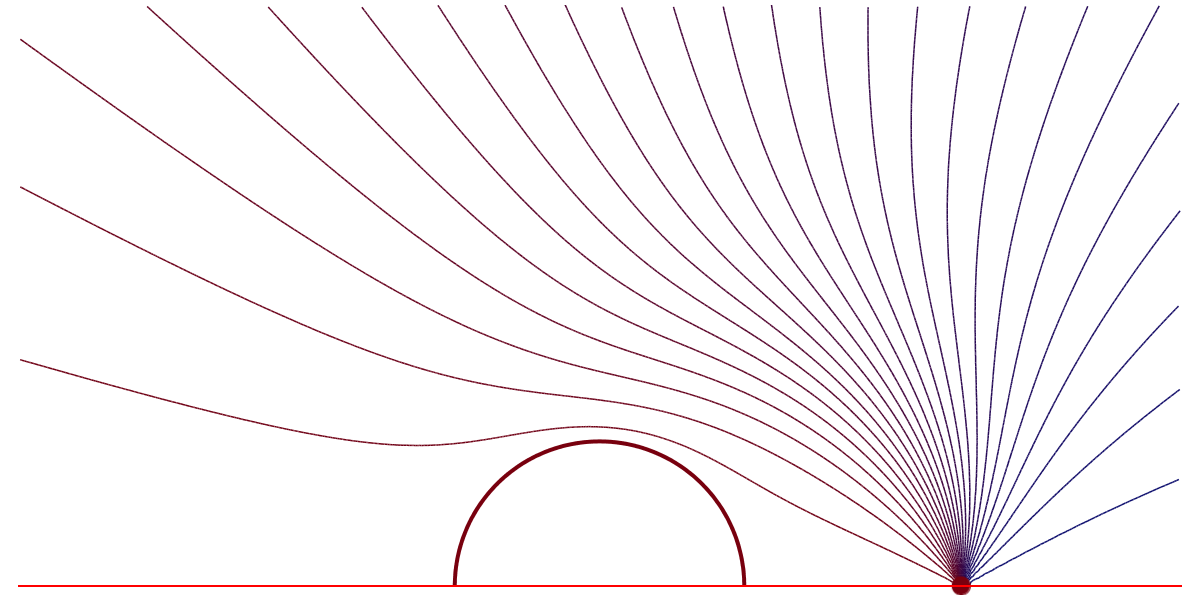}}\\
\subfloat[$\Ka=1$]{\includegraphics[width=0.32\obrA,keepaspectratio]{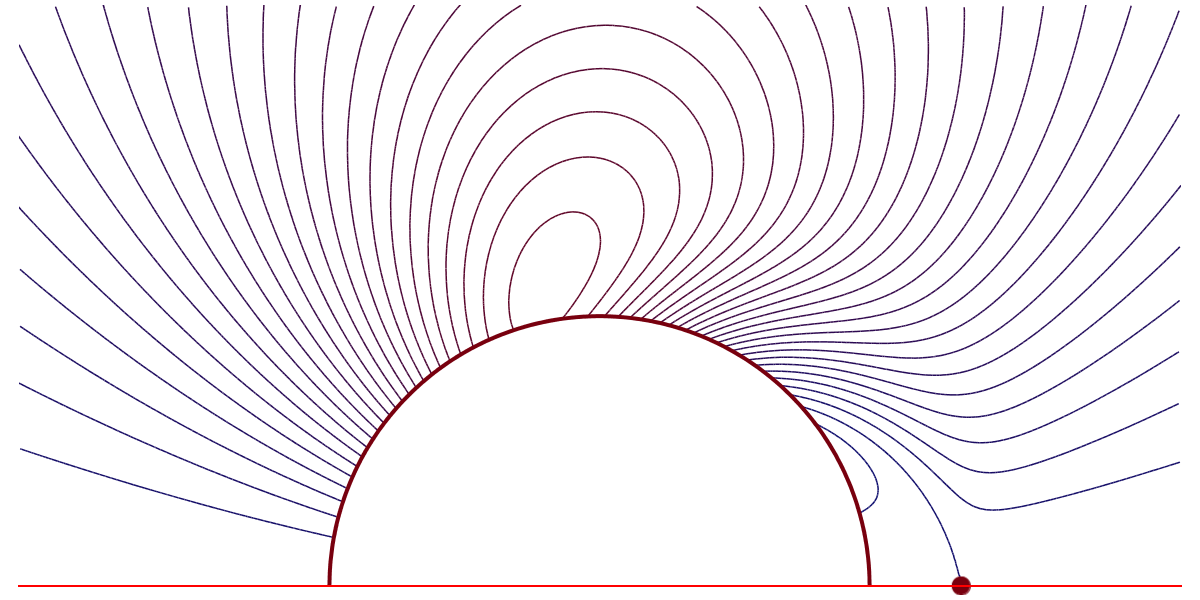}}\qquad
\subfloat[$\Ka=1.9$]{\includegraphics[width=0.32\obrA,keepaspectratio]{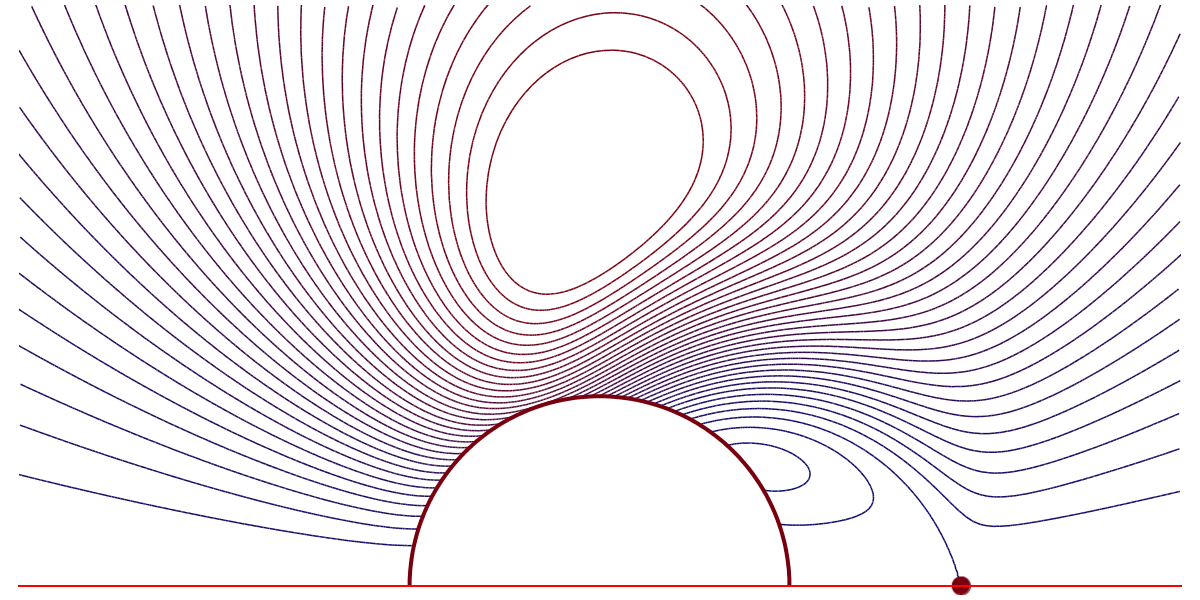}}\qquad
\subfloat[$\Ka=2$ (extremal Kerr)]{\includegraphics[width=0.32\obrA,keepaspectratio]{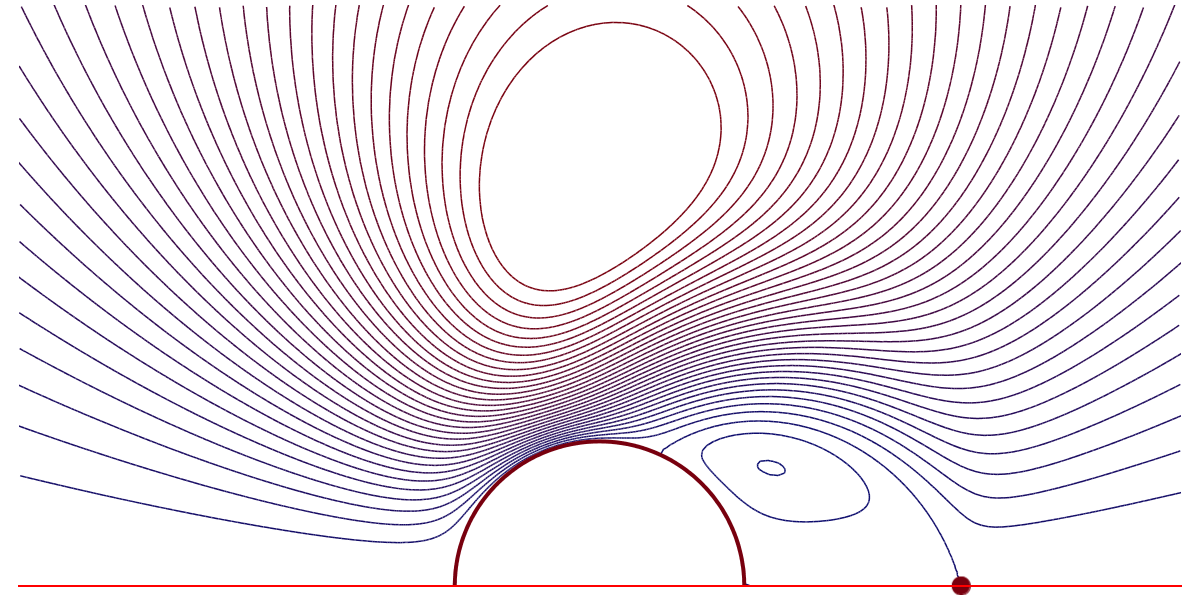}}
\caption{The electric and magnetic flux equipotentials in the first and second  second row, respectively, of a point particle at $a=5$ for the Kerr black hole of $M=2$ with varying $\Ka$. The outer black hole horizon is depicted by the thick red circle, the point particle itself by the red dot and the red horizontal line is the axis of rotation. Since the field is axially symmetric we omit the second half of the pictures.}
\vspace{1em}
\label{fig:point}
\end{figure*}

\subsection{Point particle}
Since the physical setup we are interested in is a point electric charge $q$ at $r=\az$ on the axis above the neutral Kerr black hole, our requirements are as follows
\begin{equation}
\begin{alignedat}{2}
		iQ_e+Q_m &= 0\,, &\qquad\text{for }r<\az\,, \\
		iQ_e+Q_m &= iq\,,&\qquad\text{for }r>\az\,.
\end{alignedat} \label{eq:req}
\end{equation}
For a point charge  we checked that the field generated from  our solution \re{DebyeK} with $s=1$ by means of \re{phi0}\,--\,\re{phi2} is proportional to the solution of \cite{Cohen1974}, or \cite{Bicak1976}, which is given in terms of series expansion. 

Yet, there is one difference, or correction: the presence of an electric point charge on the axis of symmetry above the Kerr black hole does not only induce an electric monopole charge on the black hole, but it also induces a \emph{magnetic} monopole charge on the black hole which has to be counterbalanced.

Taking the conditions \re{req} into account leads us to the solution
\begin{equation}
\bar{\hat{\psi}}=\frac{q}{\az^2+\Ka^2}
\left[  \left(\az-i\Ka\right) \frac{\sqrt{\xi_\Ka}}{\sin\theta} -M\left( \az\,\frac{\cos\theta}{\sin\theta}+\frac{r}{\sin\theta}\right) \right] 
\label{eq:PointCharge}\end{equation}
for the Debye potential which includes the monopole solution.

The monopole contribution corresponds to the field
\begin{equation}
2\phi_1 = qM \,\frac{i\Ka-\az}{\az^2+\Ka^2}\,\frac{1}{\left( r-i\Ka\cos\theta \right)^2}\,,
\label{eq:}\end{equation}
which would on its own represent electric and magnetic charges on the Kerr black hole given by
\begin{equation}
iQ_e+Q_m = \frac{i\az-\Ka}{\az^2+\Ka^2}\,qM\,.
\label{eq:}\end{equation}
We thus see that we had to add also a magnetic monopole proportional to the rotation parameter $\Ka$ in order to counterbalance the magnetic monopole induced on the black hole by the presence of a point (purely) electric charge on the axis. 
\begin{figure*}[t!]
\centering
\subfloat[$\Ka=1$]{\includegraphics[width=0.32\obrA,keepaspectratio]{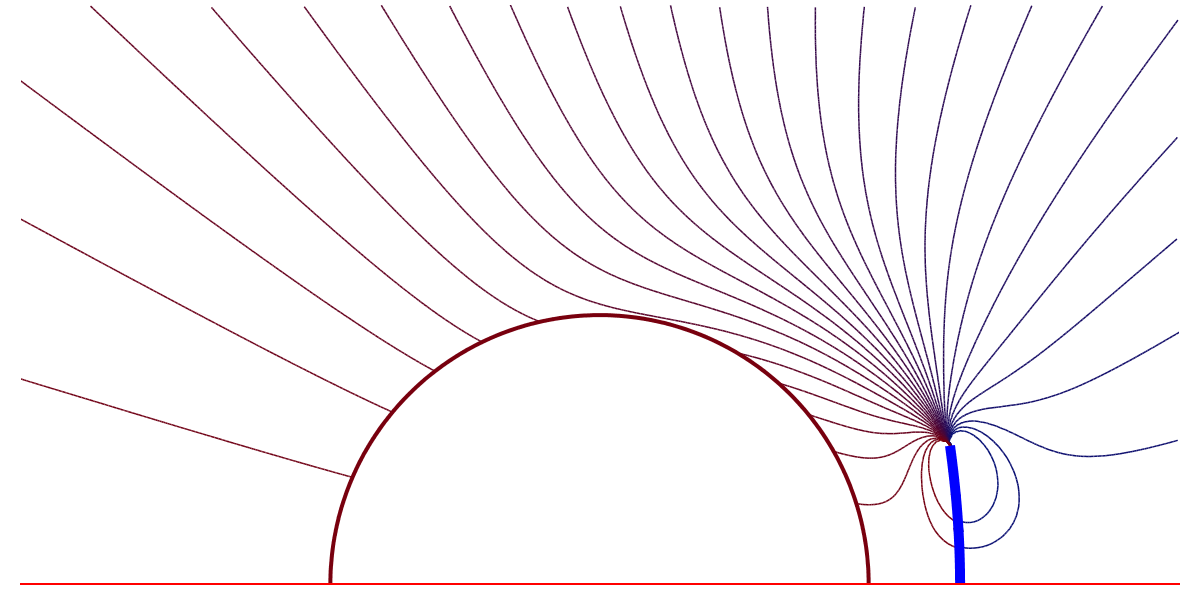}}\qquad
\subfloat[$\Ka=1.9$]{\includegraphics[width=0.32\obrA,keepaspectratio]{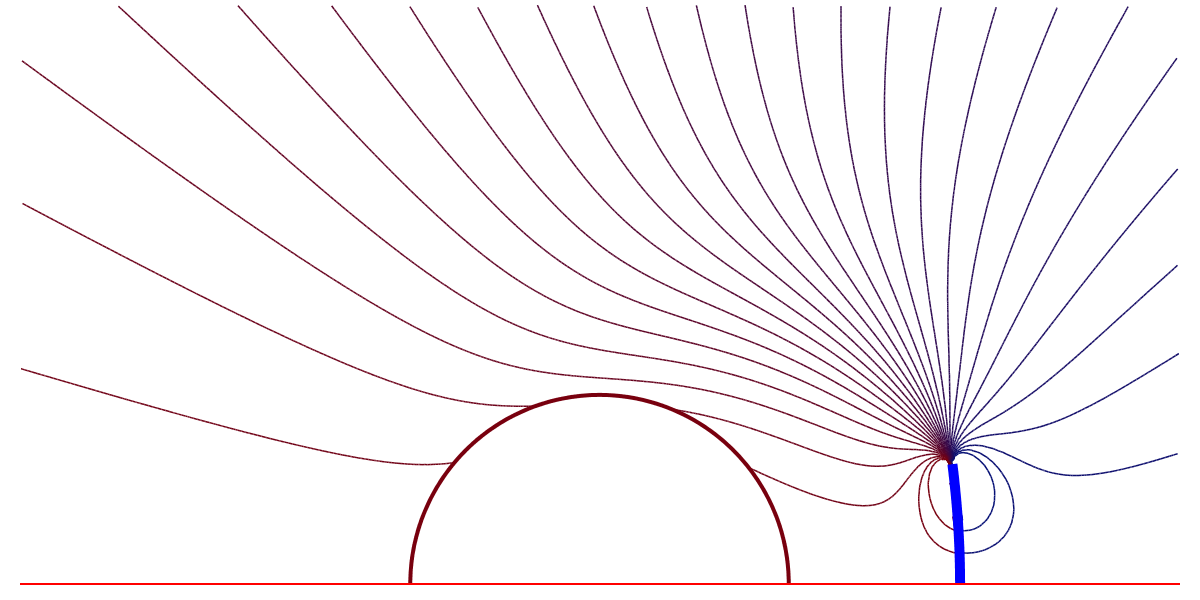}}\qquad
\subfloat[$\Ka=2$ (extremal Kerr)]{\includegraphics[width=0.32\obrA,keepaspectratio]{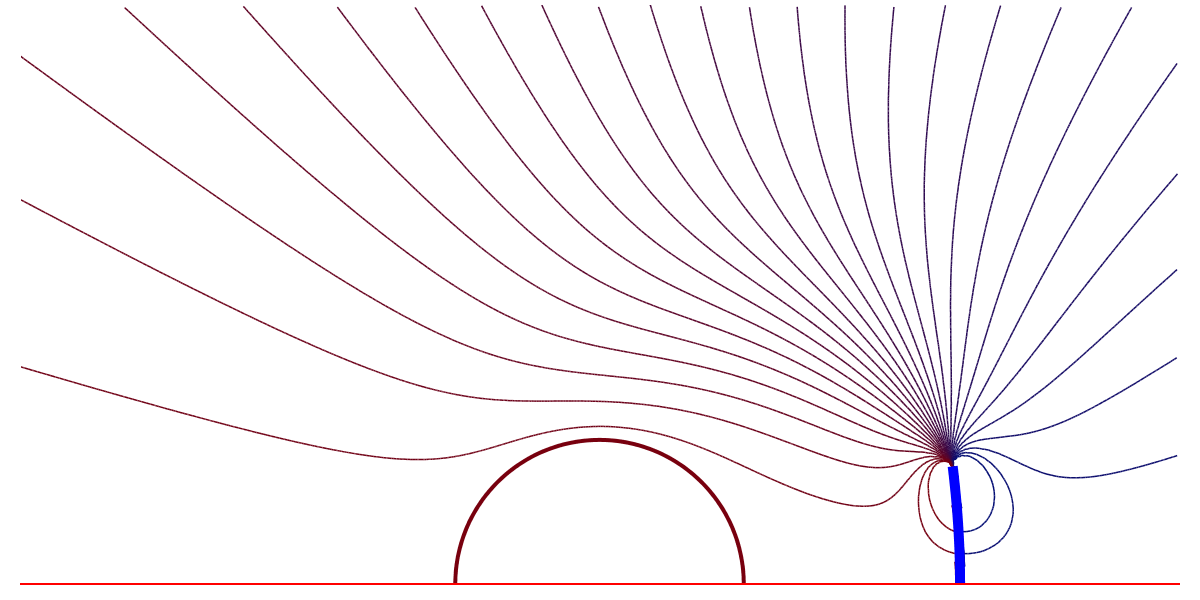}}\\
\subfloat[$\Ka=1$]{\includegraphics[width=0.32\obrA,keepaspectratio]{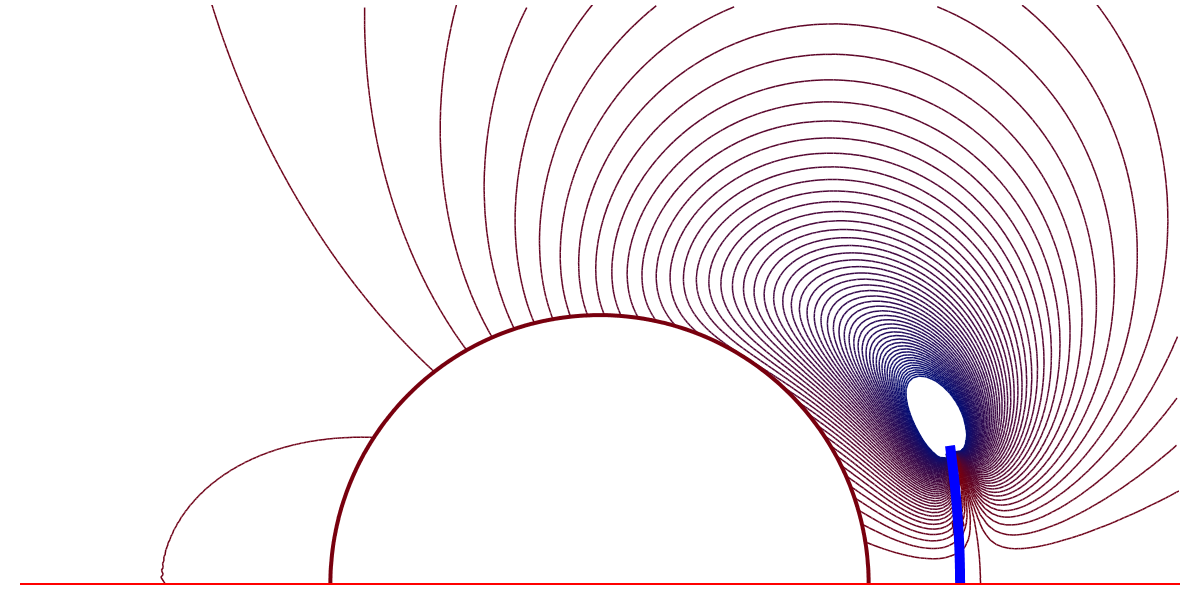}}\qquad
\subfloat[$\Ka=1.9$]{\includegraphics[width=0.32\obrA,keepaspectratio]{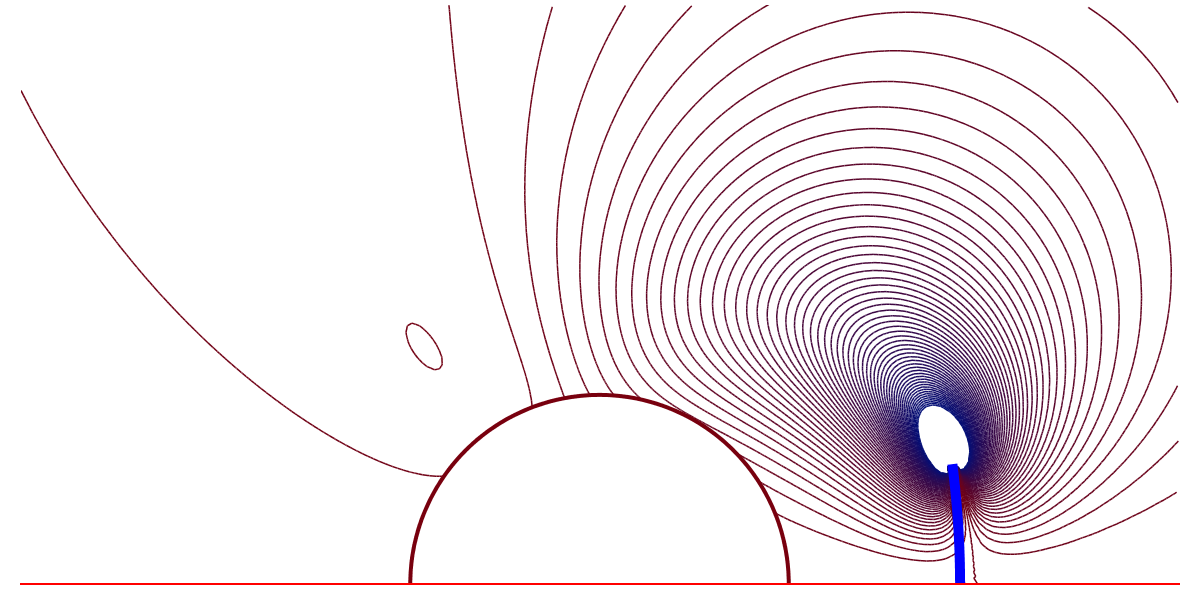}}\qquad
\subfloat[$\Ka=2$ (extremal Kerr)]{\includegraphics[width=0.32\obrA,keepaspectratio]{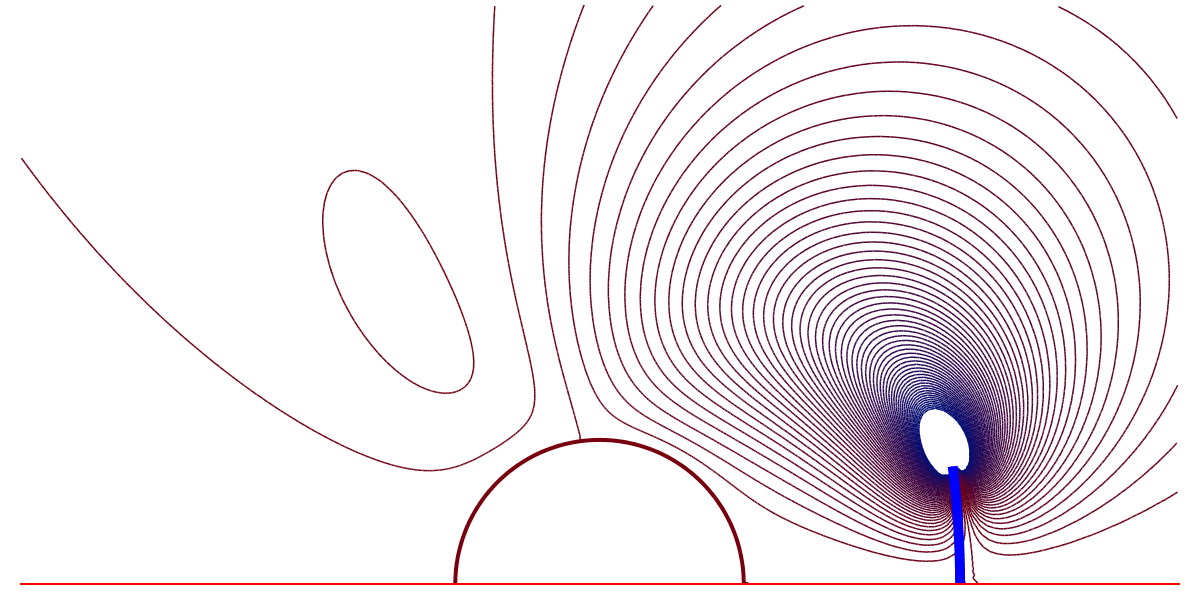}}
\caption{The electric and magnetic flux equipotentials in the first and second row, respectively, of an electromagnetic magic field at $\az=5+i$ for the Kerr black hole of $M=2$ with varying $\Ka$. The disk itself is depicted by the thick blue curve given by Eq. \re{MagicDisk}. Otherwise the legend is as in Fig.~\ref{fig:point}}
\label{fig:magic}
\end{figure*}

Analyzing the asymptotic field, we have to correct also the magnetic dipole moment of this system as given in \cite{Cohen1974} to 
\begin{equation}
\mu_{\text{mag}} = -2\,\frac{\az\Ka M q}{\az^2+\Ka^2}\,.
\label{eq:}\end{equation}

This particular solution is regular everywhere except the location of the particle itself as can be seen
\begin{itemize}
\item from the expansion in the basis given by spheroidal harmonics and solution of the Teukolsky radial equation. This expansion coincides with the solution given in \cite{Cohen1974} and is regular at the horizon,
\item from inspecting the tetrad components (ZAMO tetrad) of the electric and magnetic field which are provided in Appendix \ref{app:ZP}. Axial components of the fields are identically zero.
\end{itemize}

\subsection{Electromagnetic magic field}
The electromagnetic magic field was introduced to the special relativity by Donald Lynden-Bell \cite{Lynden-Bell2004}. It is a generalization of an old Appell's solution \cite{Appell1887} who found that it is possible to have the position of point-like particle shifted to a complex space when solving the Laplace equation.

Lynden-Bell showed that an electric charge at position $\vec{r}_0=i \Ka \vec{e_z}$ produces the same electromagnetic field as is the $G\rightarrow 0$ limit of the Kerr\,--\,Newman solution. The electromagnetic magic field is produced by a rigid rotating disc with a charge density distribution 
\begin{equation}
\sigma = -\frac{q\Ka}{2\pi\left( \Ka^2-r^2 \right)^{\nicefrac{3}{2}}}\,,
\label{eq:}\end{equation}
which is divergent but is rectified by a thin rim of opposite charge at the very end of the disc which rotates with the speed of light. We found its generalization to uniformly accelerated case, i.e. the $G\rightarrow 0$ limit of the C-metric, in \cite{Bicak2009}.

The Debye potential for a point-like particle \re{PointCharge} and \re{DebPotSchw} has its location encoded in parameters $\az$ and $\az,\,\theta_0,\,\phi_0$, respectively. There is no restriction for these to be real. Making them complex leads to an electromagnetic magic field on the Kerr and  Schwarzschild backgrounds, respectively. 

We will discuss the first case here. Substituting $\az\rightarrow \az+ir_I$ to $\xi_\Ka$ gives us
\begin{multline}
\xi_\Ka = \left(\az+ir_I-M\right)^2+\left(r-M\right)^2 \\ -2\left(\az+ir_I-M\right)\left(r-M\right)\cos\theta \\-\sin^2\theta\left(M^2-\Ka^2\right)\,.
\label{eq:cxi}\end{multline}
The position of the disc is given by the root of $\Im (\xi_\Ka)=0$, i.e. by
\begin{equation}
r_d = \frac{\az-\left(1-\cos\theta  \right)M}{\cos\theta}\,,
\label{eq:MagicDisk}\end{equation}
whereas the very end of it, the rim itself, is given by the root of $\xi_\Ka=0$ in the complex plane. The appropriate interval $\langle 0,\,\theta_d\rangle$ is bounded by
\begin{equation}
\theta_d = \arccos\left( \sqrt{\frac{\zeta-\sqrt{\zeta^2-4\left(M^2-\Ka^2\right)\left(M-\az\right)^2}}{2\left( M^2-\Ka^2 \right)}} \right),
\label{eq:}
\end{equation}
where $\zeta=M^2-\Ka^2+r_I^2+\left( M-\az \right)^2$. By computing the electromagnetic field from the Debye potential \re{PointCharge} with $\xi_\Ka$ as in \re{cxi} we can produce a field plots as in Fig.~\ref{fig:magic}. 

\subsection{Flux tubes}
We will visualise the electromagnetic field of sources hovering above the Kerr solution on the symmetry axis using the technique developed by Ruffini and Hanni \cite{Hanni1973} as a continuation of \cite{BlackHoles} and used also in \cite{Bicak2000}. This method shows the axially symmetric tubes of constant flux of electric (imaginary part) or magnetic field (real part), i.e. equipotentials in $(r,\,\theta)$ plane of the function   
\begin{equation}
i\Phi_e+\Phi_m = \frac{1}{2}\,\int_0^\theta F^*_{\theta\phi}\, \d \tilde{\theta}\,.
\label{eq:}\end{equation}

These so called ``generalized field lines'' for a point particle are shown in Fig.~\ref{fig:point}. The visualisation of the field for electromagnetic magic field on the Kerr background is in Fig.~\ref{fig:magic}.

The rotation of the spacetime induces a magnetic field. We can observe how this magnetic field is expelled out of the outer horizon as the Kerr black hole approaches extremality. This is a particular manifestation of the general phenomenon called Meissner effect.

\section{Conclusions}
We discussed a form simplification of the Teukolsky master equation (and Fackerell\,--\,Ipser equation) and shoved that the equation for the Debye potential is almost identical. A new form of Teukolsky\,--\,Starobinsky identities was presented.

Then we applied our results to point particles on the Kerr background and showed how, in a systematic way, the lowest multipoles can arise from the Debye potential. Yet, we primarily focused on compact form of solutions; not series solutions.

For an electromagnetic field we showed that a magnetic monopole must be added to a black hole in order to compensate the magnetic monopole induced by the mutual interplay of the rotation and the point charge on the axis. 

We also showed how an electromagnetic magic field generalization of a point particle is easy to obtain and plotted flux equipotentials for several configurations.

\begin{acknowledgements}
		D.K. acknowledges the support from the Czech Science Foundation, Grant No. GACR 17-16260Y. Moreover, D.K. is deeply indebted to an anonymous referee who selflessly helped to improve the manuscript. D.K. would like to thank to V. Mikeska and R. \v{S}varc for reading the manuscript.
\end{acknowledgements}

\appendix
\section{ZAMO tetrad and electromagnetic field projections}\label{app:ZP}
In order to provide a physical interpretation of the fields an observer has to be introduced. Simultaneously the fields have to be projected on an appropriate orthonormal tetrad. One of the most useful congruences of observers are zero angular momentum observers (ZAMO) whose four velocity is defined by $\vec{u}_a\propto (\vec{d} t)_a$, the congruence is thus non-twisting and as its name suggest angular momentum of particular observer vanishes, i.e. $L\equiv\vec{\eta}\cdot\vec{u}=0$. The tetrad ($\vec{u}\equiv\vec{e}_{(t)}$) is given by
\begin{equation}
\begin{alignedat}{2}
\vec{e}_{(t)} &= \frac{1}{N} \left( \vec{\p_t}+\omega\,\vec{\p_\phi} \right),\quad &
\vec{e}_{(r)} &= \sqrt{\frac{\Delta}{\Sigma}}\,\vec{\p_r}\,, \\
\vec{e}_{(\theta)} &= \frac{1}{\sqrt{\Sigma}}\,\vec{\p_\theta}\,,&
\vec{e}_{(\phi)} &= \frac{1}{\sin\theta}\,\sqrt{\frac{\Sigma}{\Upsilon}}\,\vec{\p_\phi}\,,
\end{alignedat}
\label{eq:ZAMO}
\end{equation}
where
\begin{align}
N &= \sqrt{ \frac{\left( \vec{\eta}\cdot\vec{\xi} \right)^2}{ \vec{\eta}\cdot\vec{\eta} }-\vec{\xi}\cdot\vec{\xi}}\,, &
\omega &=-\frac{\vec{\xi}\cdot\vec{\eta}}{\vec{\eta}\cdot\vec{\eta}}\,,
\end{align}
and
\begin{equation}
\Upsilon = \Delta\Sigma+2Mr\left( a^2+r^2 \right).
\label{eq:Upsilon}\end{equation}
We denote scalar product of vector $\vec{u}\cdot\vec{v}=\vec{u}^a\vec{g}_{ab}\vec{v}^b$. The Killing vectors of the Kerr metric are $\vec{\xi}=\vec{\p_t}$ and $\vec{\eta}=\vec{\p_\phi}$.
The projections 
\begin{equation}
\mathcal{E}_{(j)} = \vec{e}_{(t)}\cdot\vec{F}^*\cdot\vec{e}_{(j)}\,,\qquad\text{for}\qquad j\in (r,\,\theta,\,\phi)\,,
\end{equation}
written in compact form for $\vec{\mathcal{E}}=\vec{E}+i\vec{B}$ are
\begin{equation}
\begin{alignedat}{1}
		\mathcal{E}_{(r)} &= \frac{\frac{-i \Ka\sin\theta\Delta}{\tilde{\rho}}\,\phi_0-2\left(r^2+\Ka^2\right)\phi_1+i\Ka\sin\theta\tilde{\rho}\,\phi_2}{\sqrt{\Upsilon}}
		  \,, \\
\mathcal{E}_{(\theta)} &= \frac{r^2+\Ka^2}{\sqrt{\Delta\Upsilon}}
	\left( \frac{\Delta}{\tilde{\rho}}\,\phi_0-\frac{2i\Ka\sin\theta\Delta}{r^2+\Ka^2}\,\phi_1-\bar{\rho}\,\phi_2\right),\\
	\mathcal{E}_{(\phi)} &= -\frac{i\sqrt{\Delta}}{\tilde{\rho}}\,\phi_0-\frac{i\tilde{\rho}}{\sqrt{\Delta}}\,\phi_2\,,
\end{alignedat}
\label{eq:projE}
\end{equation}
where $\tilde{\rho}$ is a shortcut for $r-i\Ka\cos\theta$.

\bibliography{kp}

\end{document}